\documentclass[aps,pra,twocolumn,10pt,showpacs,superscriptaddress]{revtex4}

\usepackage{amsmath}
\usepackage{graphicx}
\usepackage{times}
\usepackage{amssymb}
\usepackage{hyperref}
\usepackage{xcolor}
\usepackage[T1]{fontenc}
\usepackage{bm}
\newcommand{\tj}[6]{ \begin{pmatrix}
       #1 & #2 & #3 \\
       #4 & #5 & #6 
      \end{pmatrix}}
     
\newcommand{\sj}[6]{ \begin{Bmatrix}
       #1 & #2 & #3 \\
       #4 & #5 & #6 
      \end{Bmatrix}}

\begin{document}
\title{Van der Waals interaction potential between Rydberg atoms near surfaces}
\author{Johannes Block}
\email{johannes.block@uni-rostock.de}
\author{Stefan Scheel}
\affiliation{Institut f\"ur Physik, Universit\"at Rostock, 
Albert-Einstein-Strasse 23, D-18059 Rostock, Germany}
\date{\today} 

\begin{abstract}
Van der Waals interactions, as a result of the exchange of photons between 
particles, can be altered by modifying the environment through which these 
photons propagate. As a consequence, phenomena such as the Rydberg blockade 
mechanism between highly excited atoms or excitons can be controlled by adding 
reflecting surfaces. We provide the quantum electrodynamic framework for the 
van der Waals interaction in the nonretarded limit that is relevant for 
long-wavelength transitions such as those between Rydberg systems, and show its 
intimate connection with common static dipole-dipole interactions, thereby 
providing a generalization to include macroscopic bodies.
\end{abstract}

\maketitle

\section{Introduction}

Dispersion forces such as Casimir or van der Waals interactions arise from 
(ground-state or thermal) fluctuations of the quantized electromagnetic field 
that spontaneously induce a (dipole) polarization in matter. The resulting 
dipole-dipole interaction, e.g. between polarizable atoms, then gives rise to a 
typically attractive force. Early investigations focussed on fluctuating 
dipoles which, after the pioneering works of H.B.G.~Casimir who pointed out the
important role of the quantized electromagnetic field, could be 
traced back to its quantum electrodynamic origins. Today, we have a solid
understanding of the origin of dispersion forces \cite{BuhmannI,Salam} and 
their role in molecular and surface science \cite{Israelachvili}. 

Van der Waals forces, i.e. dispersion forces between atomic or molecular
systems, are long-range interactions when compared to all electronic 
interactions such as the Pauli repulsion. However, as the van der Waals 
potential decays as $r^{-6}$ at close separation, and as $r^{-7}$ for larger 
distances, their effect is typically only important for (sub-)nanometer 
distances. An exception is the van der Waals interaction between highly excited 
(Rydberg) atoms that, due to the scaling of the $C_6$ coefficient as $n^{11}$ 
with the principal quantum number $n$, can extend over several micrometers and 
cause the Rydberg blockade effect \cite{Lukin01}. There, an atomic line is 
shifted out of resonance by the presence of a Rydberg atom nearby. This effect 
is the basis for a large number of quantum-optical protocols including the 
deterministic generation of single photons on demand \cite{Dudin12} and the 
demonstration of a single-photon transistor \cite{Tiarks14}.

The van der Waals interaction potential of ground-state atoms and molecules in 
the presence of arbitrary magnetodielectric bodies is well understood, and can 
be formulated within the framework of macroscopic quantum electrodynamics 
\cite{Acta} as a fourth-order perturbation with respect to the single-atom 
electric-dipole interaction Hamiltonian \cite{Safari06}. For excited atoms or 
molecules, the shape of the interaction potential is known in the nonretarded 
limit, but controversial discussions remain as to its behaviour in the retarded 
limit \cite{Gomberoff66,Power95,Safari15,Donaire15,Barcellona16}, whether the 
interaction remains monotonous with distance or whether is oscillates as 
expected from equivalent Casimir--Polder results \cite{Ellingsen10,Ellingsen11}.

In the nonretarded limit, i.e. when the separation between the atoms or 
molecules is much smaller than all wavelengths associated with optical 
transitions in the atoms and the surrounding macroscopic bodies, the van der 
Waals interaction has to be taken in the static (low-frequency) limit. It is 
then typically computed in second-order perturbation theory with respect to the 
static dipole-dipole interaction Hamiltonian \cite{Saffman08}. 

In this article, we will show how the static limit of the van der Waals
interaction for excited atoms follows from the general fourth-order perturbation
theory and, in particular, note that the distinction between nonresonant and
resonant contributions to the interaction that is usually made in this context
is misleading here. This derivation provides a useful generalization of the 
van der Waals interaction between highly excited (Rydberg) atoms near 
(magneto-)dielectric surfaces. As an application, we will show how the presence 
of boundaries, for example metallic structures for generating atom traps or
confining geometries such as thin crystals, influences the blockade radius 
between Rydberg atoms and excitons.

The article is organized as follows: In Sec.~\ref{sec:vdWnonretarded} we derive 
the nonretarded limit of the van der Waals interaction potential using the 
formalism of macroscopic quantum electrodynamics and show how nonresonant and 
resonant contributions combine to add up to a generalized version of the static 
interaction that now also holds for arbitrary geometries. In 
Sec.~\ref{sec:confined} we provide some examples of the effects of confining 
geometries on the van der Waals interaction between excited atoms and excitons. 
In Sec.~\ref{sec:atoms}, we provide a numerical example of body-assisted van 
der Waals interaction of a pair of Rydberg atoms near a perfectly conducting 
mirror and discuss its influence on the Rydberg blockade mechanism. 

\section{Van der Waals interaction in the nonretarded limit}
\label{sec:vdWnonretarded}

The van der Waals interaction potential between two atoms prepared in energy 
eigenstates $|k_A\rangle$ and $|l_B\rangle$, respectively, can be derived from 
fourth-order perturbation theory as \cite{Safari15}
\begin{gather}
U_{AB}(\mathbf{r}_A,\mathbf{r}_B) = \frac{i\mu_0^2}{\hbar\pi} 
\sum\limits_{m\ne k,n\ne l} \frac{1}{\omega_A^{mk}+\omega_B^{nl}} \nonumber\\
\times \bigg\{ \mathcal{P}\int\limits_0^\infty d\omega 
\frac{\omega^4(\omega_A^{mk}+\omega_B^{nl}+\omega)}
{(\omega+\omega_A^{mk})(\omega+\omega_B^{nl})} \nonumber \\
+\mathcal{P}\int\limits_0^{-\infty} d\omega 
\frac{\omega^4(\omega_A^{mk}+\omega_B^{nl}-\omega)}
{(\omega-\omega_A^{mk})(\omega-\omega_B^{nl})} \bigg\} \nonumber \\
\times \left[ \mathbf{d}_A^{km} \cdot \bm{G}(\mathbf{r}_A,\mathbf{r}_B,\omega) 
\cdot \mathbf{d}_B^{ln} \right]^2
\label{eq:vdWgeneral}
\end{gather}
where $\mathbf{d}_A^{km}=\langle k_A|\hat{\mathbf{d}}_A|m_A\rangle$ and 
$\omega_A^{km}$ denote the electric dipole moments and the transition 
frequencies. All information about geometric and optical properties of 
macroscopic bodies surrounding the atoms is contained in the dyadic Green 
function $\bm{G}(\mathbf{r}_A,\mathbf{r}_B,\omega)$ as fundamental solution of 
the vector Helmholtz equation
\begin{equation}
\bm{\nabla}\times\bm{\nabla}\times\bm{G}(\mathbf{r},\mathbf{r}',\omega) 
-\varepsilon(\mathbf{r},\omega) \frac{\omega^2}{c^2} 
\bm{G}(\mathbf{r},\mathbf{r}',\omega) = \delta(\mathbf{r}-\mathbf{r}') \bm{I}.
\end{equation}

If one or both atoms is/are initially prepared in an excited state, the 
integrand in Eq.~(\ref{eq:vdWgeneral}) has poles on the real axis. Hence, using 
contour integration techniques, one can transform the integral along the real 
axis into an integral over imaginary frequencies and a sum over the residues at 
the poles on the real axis. In this way, a distinction into contributions from 
nonresonant (integral along imaginary axis) and resonant (residues) 
interactions can be made \cite{Buhmann04}, that can partially cancel one 
another \cite{Ellingsen10,Ellingsen11}. These contributions are commonly 
discussed separately due to their apparently different origins. However, we 
will show that, in particular instances relevant to Rydberg physics, such 
distinction is spurious and unhelpful. Moreover, it disguises the vital 
connection between static and dynamic van der Waals interactions.

The physics of interactions between highly excited (Rydberg) atoms is dominated 
by nonretarded, i.e. static interactions due to the long transition wavelengths 
between neighboring Rydberg states. In the nonretarded limit, we can take the 
zero-frequency limit of the dyadic Green function (denoted $\bm{\Gamma}_0$ in 
Ref.~\cite{Ellingsen11}) defined as 
\begin{equation}
\bm{\Gamma}_0(\mathbf{r}_A,\mathbf{r}_B) = \lim\limits_{\omega\to0} 
\frac{\omega^2}{c^2} \bm{G}(\mathbf{r}_A,\mathbf{r}_B,\omega).
\end{equation}
It exists for all Green functions due to their general property \cite{Acta}
\begin{equation}
\operatorname{Re}\bm{G}(\mathbf{r},\mathbf{r}',\omega) 
\stackrel{|\omega|\to0}{\sim} \omega^{-2},\quad
\operatorname{Im}\bm{G}(\mathbf{r},\mathbf{r}',\omega) 
\stackrel{|\omega|\to0}{\sim} \omega^{-1}.
\end{equation}
Hence, in the nonretarded limit, the van der Waals interaction potential 
(\ref{eq:vdWgeneral}) reads as
\begin{gather}
U_{AB}(\mathbf{r}_A,\mathbf{r}_B) = \frac{i}{\hbar\pi\varepsilon_0^2} 
\sum\limits_{m\ne k,n\ne l} \frac{\left[ \mathbf{d}_A^{km} \cdot 
\bm{\Gamma}_0(\mathbf{r}_A,\mathbf{r}_B) \cdot \mathbf{d}_B^{ln} 
\right]^2}{\omega_A^{mk}+\omega_B^{nl}} \nonumber\\
\times \bigg\{ \mathcal{P}\int\limits_0^\infty d\omega 
\frac{\omega_A^{mk}+\omega_B^{nl}+\omega}
{(\omega+\omega_A^{mk})(\omega+\omega_B^{nl})} \nonumber \\
+\mathcal{P}\int\limits_0^{-\infty} d\omega 
\frac{\omega_A^{mk}+\omega_B^{nl}-\omega}
{(\omega-\omega_A^{mk})(\omega-\omega_B^{nl})} \bigg\}
\label{eq:vdWnonretarded}
\end{gather}
in which spatial dependence and frequency-dependence are decoupled and can be 
treated separately.

Let us focus on the frequency integral in Eq.~(\ref{eq:vdWnonretarded}) first. 
The double sum over all available atomic transitions determines whether 
resonant contributions to the van der Waals interaction potential exist. In 
order to determine the individual contributions of upward and downward 
transitions, we split the double sum into four distinct contributions
\begin{equation}
\sum\limits_{m\ne k,n\ne l} \equiv \sum\limits_{m>k,n>l} + 
\sum\limits_{m<k,n>l} + \sum\limits_{m>k,n<l} + \sum\limits_{m<k,n<l}.
\end{equation}

We begin with the situation that both atoms perform upward transitions 
($m>k,n>l$) in which case the transition frequencies 
$\omega_A^{mk}$ and $\omega_B^{nl}$ are both strictly positive. Hence, the 
integrals along the real $\omega$ axis can be flipped to the imaginary axis, 
noting that the contributions from the two quarter-circles cancel one another. 
We thus find
\begin{gather}
\mathcal{P}\int\limits_0^\infty d\omega 
\frac{\omega_A^{mk}+\omega_B^{nl}+\omega}
{(\omega+\omega_A^{mk})(\omega+\omega_B^{nl})} \nonumber\\
= i\int\limits_0^\infty d\xi 
\frac{\omega_A^{mk}+\omega_B^{nl}+i\xi}
{(i\xi+\omega_A^{mk})(i\xi+\omega_B^{nl})},\\
\mathcal{P}\int\limits_0^{-\infty} d\omega 
\frac{\omega_A^{mk}+\omega_B^{nl}-\omega}
{(\omega-\omega_A^{mk})(\omega-\omega_B^{nl})} \nonumber\\
= i\int\limits_0^\infty d\xi 
\frac{\omega_A^{mk}+\omega_B^{nl}-i\xi}
{(i\xi-\omega_A^{mk})(i\xi-\omega_B^{nl})}.
\end{gather}
The integrations can be performed after adding both integrals with the result 
that
\begin{gather}
\mathcal{P}\int\limits_0^\infty d\omega 
\frac{\omega_A^{mk}+\omega_B^{nl}+\omega}
{(\omega+\omega_A^{mk})(\omega+\omega_B^{nl})} \nonumber\\
+\mathcal{P}\int\limits_0^{-\infty} d\omega 
\frac{\omega_A^{mk}+\omega_B^{nl}-\omega}
{(\omega-\omega_A^{mk})(\omega-\omega_B^{nl})} = i \pi.
\end{gather}

In contrast, let us investigate the contribution to the integral from downward 
transitions ($m<k,n<l$). Here we write
\begin{gather}
\mathcal{P}\int\limits_0^\infty d\omega 
\frac{\omega_A^{mk}+\omega_B^{nl}+\omega}
{(\omega+\omega_A^{mk})(\omega+\omega_B^{nl})} \nonumber\\
= -i\int\limits_0^\infty d\xi 
\frac{\omega_A^{km}+\omega_B^{ln}-i\xi}
{(i\xi-\omega_A^{km})(i\xi-\omega_B^{ln})} +i\pi,\\
\mathcal{P}\int\limits_0^{-\infty} d\omega 
\frac{\omega_A^{mk}+\omega_B^{nl}-\omega}
{(\omega-\omega_A^{mk})(\omega-\omega_B^{nl})} \nonumber\\
= -i\int\limits_0^\infty d\xi 
\frac{\omega_A^{km}+\omega_B^{ln}+i\xi}
{(i\xi+\omega_A^{km})(i\xi+\omega_B^{ln})} +i\pi,
\end{gather}
where the pole contributions at $\omega=\omega_A^{km}$ and 
$\omega=\omega_B^{ln}$ have been added to the nonresonant contributions. Adding 
all up, we find that
\begin{gather}
\mathcal{P}\int\limits_0^\infty d\omega 
\frac{\omega_A^{mk}+\omega_B^{nl}+\omega}
{(\omega+\omega_A^{mk})(\omega+\omega_B^{nl})} \nonumber\\
+\mathcal{P}\int\limits_0^{-\infty} d\omega 
\frac{\omega_A^{mk}+\omega_B^{nl}-\omega}
{(\omega-\omega_A^{mk})(\omega-\omega_B^{nl})} = -i\pi +2i\pi =i\pi
\end{gather}
where the resonant contributions ($=2i\pi$) overcompensate the nonresonant ones 
($=-i\pi$).

Finally, the contributions to the double sum when one atom performs an upward 
transition and the other a downward transition ($m<k,n>l$ and $m>k,n<l$) can be 
written as
\begin{gather}
\mathcal{P}\int\limits_0^\infty d\omega 
\frac{\omega_A^{mk}+\omega_B^{nl}+\omega}
{(\omega+\omega_A^{mk})(\omega+\omega_B^{nl})} \nonumber\\
+\mathcal{P}\int\limits_0^{-\infty} d\omega 
\frac{\omega_A^{mk}+\omega_B^{nl}-\omega}
{(\omega-\omega_A^{mk})(\omega-\omega_B^{nl})} \nonumber\\
= i\pi \frac{\omega_A^{km}-\omega_B^{nl}}{\omega_A^{km}+\omega_B^{nl}} 
+2i\pi\frac{\omega_B^{nl}}{\omega_A^{km}+\omega_B^{nl}} =i\pi
\end{gather}
($m<k,n>l$) and
\begin{gather}
\mathcal{P}\int\limits_0^\infty d\omega 
\frac{\omega_A^{mk}+\omega_B^{nl}+\omega}
{(\omega+\omega_A^{mk})(\omega+\omega_B^{nl})} \nonumber\\
+\mathcal{P}\int\limits_0^{-\infty} d\omega 
\frac{\omega_A^{mk}+\omega_B^{nl}-\omega}
{(\omega-\omega_A^{mk})(\omega-\omega_B^{nl})} \nonumber\\
= -i\pi \frac{\omega_A^{mk}-\omega_B^{ln}}{\omega_A^{mk}+\omega_B^{ln}} 
+2i\pi\frac{\omega_A^{mk}}{\omega_A^{mk}+\omega_B^{ln}} =i\pi
\end{gather}
($m>k,n<l$), where we highlighted both nonresonant and resonant contributions.

Intriguingly, despite the fact that the frequency integrals along the real axis 
may or may not contain poles depending on the nature of the atomic transition, 
the sum of nonresonant and resonant contributions to the integral always give 
identical results, with the pole contributions (over-)compensating the 
integrations along the imaginary axis. As a result, at least in the nonretarded 
limit in which frequency- and spatial dependence of the interaction potential 
separate, the split into nonresonant and resonant contributions is artificial 
and not helpful. The temperature invariance of Casimir--Polder forces in a 
similarly limiting regime \cite{Ellingsen10,Ellingsen11} is a consequence of 
the same type of arguments. 

We are thus left with the van der Waals interaction 
potential in the nonretarded limit as
\begin{equation}
U_{AB}(\mathbf{r}_A,\mathbf{r}_B) = -\frac{1}{\hbar\varepsilon_0^2} 
\sum\limits_{m\ne k,n\ne l} \frac{\left[ \mathbf{d}_A^{km} \cdot 
\bm{\Gamma}_0(\mathbf{r}_A,\mathbf{r}_B) \cdot \mathbf{d}_B^{ln} 
\right]^2}{\omega_A^{mk}+\omega_B^{nl}}.
\label{eq:vdWstatic}
\end{equation}
Equation~(\ref{eq:vdWstatic}) can equivalently be obtained from second-order 
perturbation theory with the dipole-dipole interaction Hamiltonian
\begin{equation}
\hat{V}_{dd} = \frac{1}{\varepsilon_0} \hat{\mathbf{d}}_A \cdot 
\bm{\Gamma}_0(\mathbf{r}_A,\mathbf{r}_B) \cdot \hat{\mathbf{d}}_B .
\end{equation}
In free space, the dyadic Green function can be constructed from the static, 
scalar Green function 
\begin{equation}
g_0(\mathbf{r},\mathbf{r}') = \frac{1}{4\pi|\mathbf{r}-\mathbf{r}'|}
\end{equation}
as
\begin{equation}
\bm{\Gamma}_0(\mathbf{r},\mathbf{r}') = \bm{\nabla}\otimes\bm{\nabla} 
g_0(\mathbf{r},\mathbf{r}') = -\frac{1}{4\pi} \left[ \bm{I} - 
3\mathbf{e}_{\bm{\varrho}}\otimes\mathbf{e}_{\bm{\varrho}} \right]
\end{equation}
with $\bm{\varrho}=\mathbf{r}-\mathbf{r}'$ and 
$\mathbf{e}_{\bm{\varrho}}=\bm{\varrho}/|\bm{\varrho}|$, so that the 
dipole-dipole interaction potential takes its familiar form
\begin{equation}
\hat{V}_{dd} = -\frac{1}{4\pi\varepsilon_0} \left[ \hat{\mathbf{d}}_A \cdot  
\hat{\mathbf{d}}_B -3(\hat{\mathbf{d}}_A \cdot\mathbf{e}_{\bm{\varrho}})  
(\hat{\mathbf{d}}_B \cdot\mathbf{e}_{\bm{\varrho}}) \right].
\end{equation}
It is precisely this form of the interaction potential which is used in atomic 
physics to describe the van der Waals interaction between highly excited 
(Rydberg) atoms \cite{Saffman08}, if one associates the energy denominator 
$\hbar(\omega_A^{mk}+\omega_B^{nl})$ in Eq.~(\ref{eq:vdWstatic}) with the 
F\"orster defect.

\section{Nonretarded van der Waals interaction in confined geometries}
\label{sec:confined}

The advantage of the Green function formulation of the nonretarded van der 
Waals interaction potential (\ref{eq:vdWstatic}) becomes obvious as soon as the 
interacting atoms are not located in free space but in the vicinity of some 
macroscopic body such as trapping structures, mirrors etc. In these situations, 
the computation of the Green function in the nonretarded or 
static limit is simplified as the well-known image dipole construction applies. 

\begin{figure}
\includegraphics[width=0.6\columnwidth]{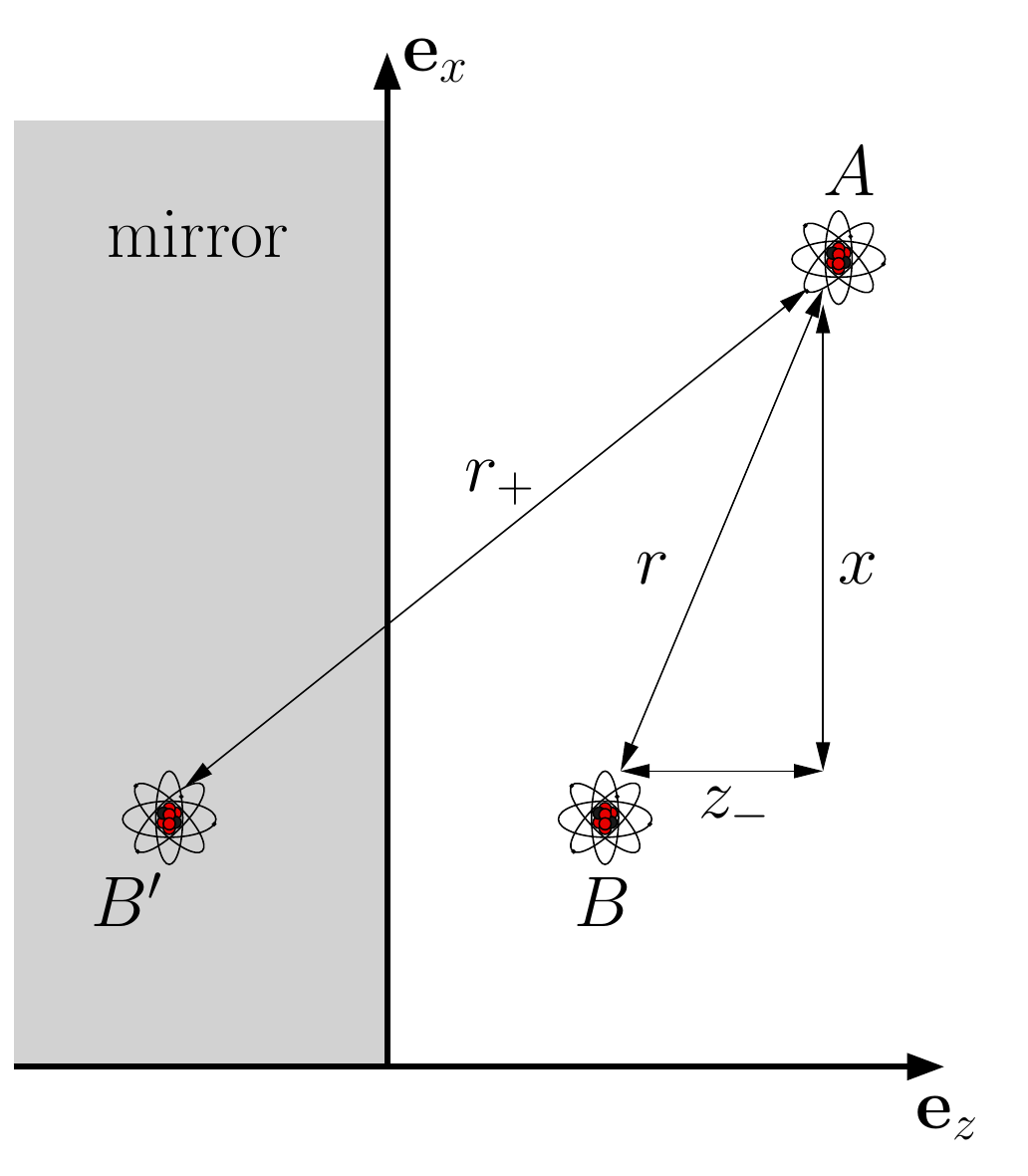}
\caption{Two atoms in front of perfectly conducting mirror. In case of parallel 
alignment we have $z_- = 0$.} 
\label{fig:2atome}
\end{figure}

As a first example, we consider two atoms in front of a perfectly reflecting 
mirror. By the method of images, the dyadic Green function can be constructed 
from the scalar Green functions 
\begin{equation}
g_0(\mathbf{r},\mathbf{r}',\omega) = 
\frac{e^{i\omega|\mathbf{r}-\mathbf{r}'|/c}}{4\pi|\mathbf{r}-\mathbf{r}'|}
\end{equation}
in real space and 
\begin{equation}
g_0(\mathbf{r},\mathbf{r}'_i,\omega) = 
\frac{e^{i\omega|\mathbf{r}-\mathbf{r}'|/c}}{4\pi|\mathbf{r}-\mathbf{r}'_i|}
\end{equation}
in reflected space with the coordinates $\mathbf{r}'_i=(x',y',-z')$ of the 
image dipole as \cite{Tai}
\begin{gather}
\bm{G}(\mathbf{r},\mathbf{r}',\omega) = \left[ \bm{I}-\frac{c^2}{\omega^2} 
\bm{\nabla}\otimes\bm{\nabla}' \right] \left[ g_0(\mathbf{r},\mathbf{r}',\omega) 
-g_0(\mathbf{r},\mathbf{r}'_i,\omega) \right] \nonumber\\
+2\mathbf{e}_z\otimes\mathbf{e}_z 
g_0(\mathbf{r},\mathbf{r}'_i,\omega).
\end{gather}
In the nonretarded (i.e. static) limit, this becomes 
\begin{gather}
\bm{\Gamma}_0(\mathbf{r}_A,\mathbf{r}_B) = -\frac{1}{4\pi} \left[ \frac{1}{r^3}
\begin{pmatrix}
1 & 0 & 0 \\ 0 & 1 & 0 \\ 0 & 0 & 1
\end{pmatrix}
-\frac{3}{r^5}
\begin{pmatrix}
x^2 & 0 & xz_- \\ 0 & 0 & 0 \\ xz_- & 0 & z_-^2
\end{pmatrix}
\right] \nonumber\\
+\frac{1}{4\pi} \left[ 
\frac{1}{r_+^3} 
\begin{pmatrix}
1 & 0 & 0 \\ 0 & 1 & 0 \\ 0 & 0 & 2
\end{pmatrix}
-\frac{3}{r_+^5}
\begin{pmatrix}
x^2 & 0 & -xz_+ \\ 0 & 0 & 0 \\ xz_+ & 0 & x^2
\end{pmatrix}
\right]
\label{eq:staticGmirror}
\end{gather}
where $x=x_A-x_B$ and $z_\pm=z_A\pm z_B$. The distance between the dipoles at 
$\mathbf{r}_A$ and $\mathbf{r}_B$ is denoted by $r$, the distance between the 
dipole at $\mathbf{r}_B$ and the image dipole at $\mathbf{r}'_{A}$ is $r_+$ (see 
Figure~\ref{fig:2atome}).

The Green function (\ref{eq:staticGmirror}) thus consists of two parts, the 
Coulomb potential of the dipole at $\mathbf{r}_B$ and the potential of the 
image dipole at $\mathbf{r}'_B$, both evaluated at the observation point 
$\mathbf{r}_A$. Taking the square of the Green function as necessary in 
Eq.~(\ref{eq:vdWstatic}) results in the usual interpretation of van der Waals 
interactions near surfaces: the interaction can be subdivided into a direct 
photon exchange, a photon exchange with the image dipoles, and a mixed 
contribution of direct and image dipole interaction.

\begin{figure}[ht]
\includegraphics[width=8cm]{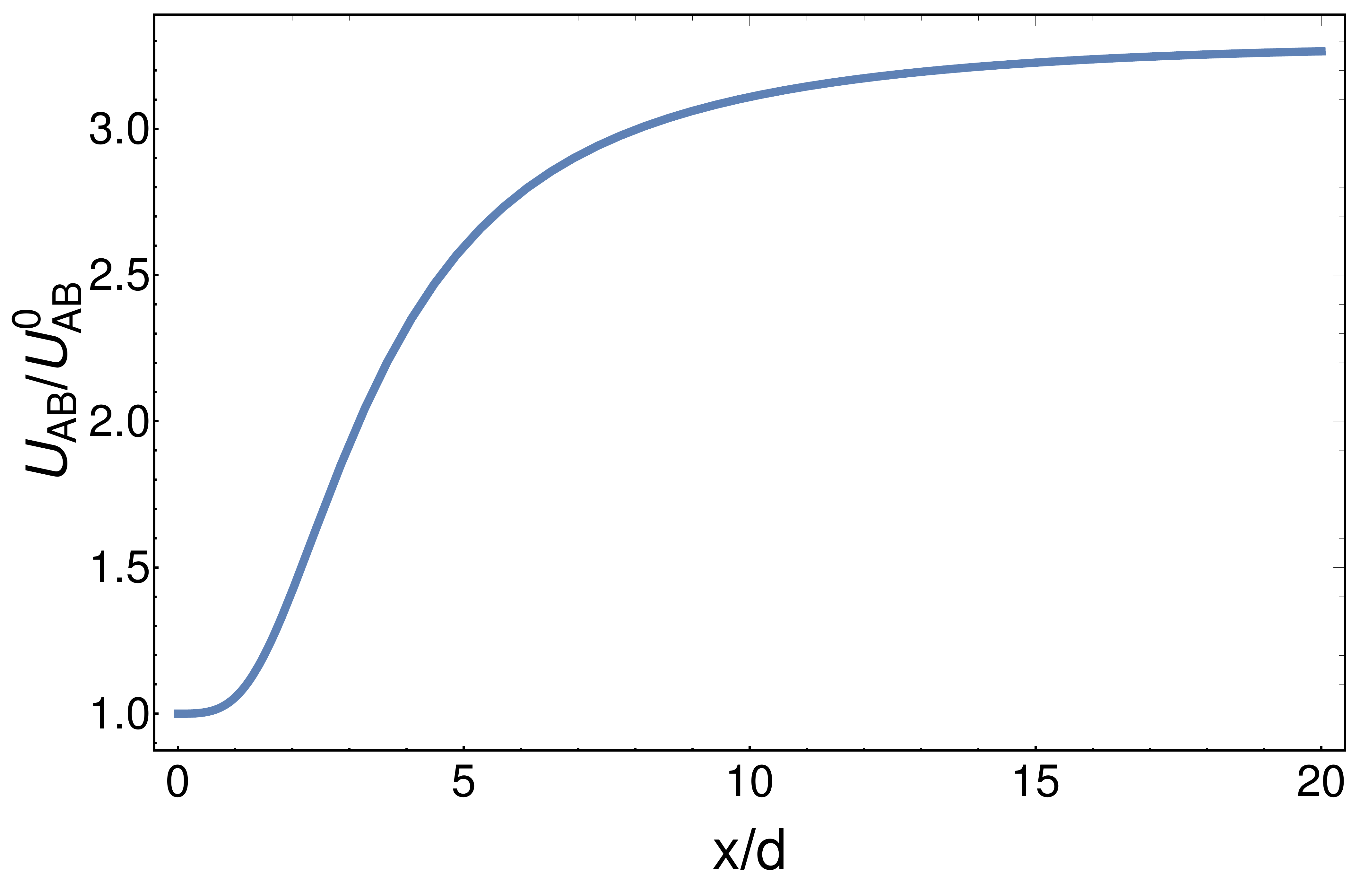}
\includegraphics[width=8cm]{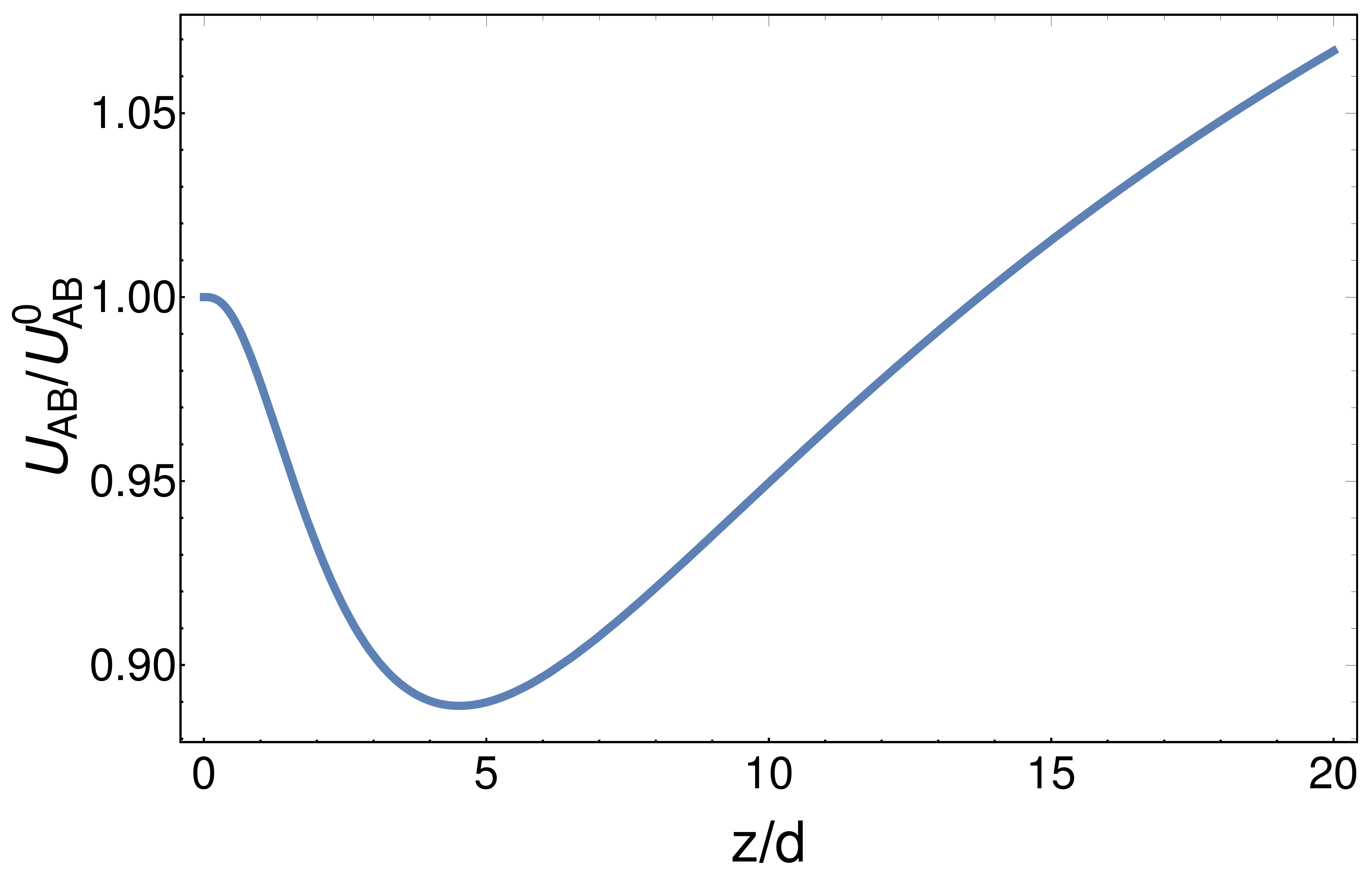}
\caption{\label{fig:static} Modified van der Waals interaction potential for 
horizontal (top) and vertical (bottom) alignment of the dimer near a perfect 
mirror, normalized with respect to the free-space value.}
\end{figure}

As a result, close to a surface, the van der Waals interaction is modified 
according to the orientation of the dimer with respect to the surface. As a 
first illustrative example, in Figure~\ref{fig:static} we show the relative 
strength of the van der Waals interaction for isotropically averaged dipoles 
near a perfectly conducting mirror.
In both figures, the distance of the closest dipole from the mirror is $d$. The 
top figure refers to a horizontally aligned dimer separated by a distance $x$, 
whereas the bottom figure depicts the interaction potential of a vertically 
aligned dimer separated by a distance $z$. One observes that the van der Waals 
interaction between horizontally aligned Rydberg dimers can exceed their 
free-space value by a factor of up to $10/3$ which has immediate consequences 
for the Rydberg blockade between atoms near surfaces as we will show later.

For real surfaces with less than unit reflectivity, the strength of the image 
dipole decreases by a factor equal to the (static) reflection coefficient 
$r_p$. The corresponding Green function can then be constructed from 
Eq.~(\ref{eq:staticGmirror}) by multiplying the second term in square brackets 
by $r_p$ or, equivalently, by taking the nonretarded limit of the Green 
function presented in Appendix~C of Ref.~\cite{Safari06}. In this case, the 
maximal enhancement factor of horizontally aligned dimers drops to
\begin{equation}
\lim\limits_{d\to0}
\frac{U_{AB}(\mathbf{r}_A,\mathbf{r}_B)}
{U^{(0)}_{AB}(\mathbf{r}_A,\mathbf{r}_B)} 
= r_p^2+\frac{4}{3}r_p+1.
\end{equation}

\subsection{Semiconductor Rydberg excitons in finite-thickness crystals}

The realm of Rydberg physics has recently been extended to include excitons in 
semiconductors \cite{Kazimierczuk14}, which provide a quasiparticle analogue to 
atoms. In this case, however, the excitons are created and thus confined in a 
crystal of finite thickness. Hence, the van der Waals interaction between the 
Rydberg excitons has to be considered inside a planar cavity (in case
of the experiment reported in Ref.~\cite{Kazimierczuk14} a crystal slab of
$34\mu$m thickness). As the dominant transitions that contribute to the van der
Waals interaction potential have millimeter wavelengths (the Rydberg energy of
the excitons of the yellow series in $\mathrm{Cu_2O}$ is around $86$meV), the 
nonretarded limit applies even for this rather thick crystal.

As the excitons are embedded in a host material, the van der Waals interaction 
is screened by the crystal environment. It has been shown that the local-field 
corrected two-point Green function is \cite{Sung06}
\begin{equation}
\bm{G}_\mathrm{loc}(\mathbf{r}_A,\mathbf{r}_B,\omega) = 
\frac{3\varepsilon_A}{2\varepsilon_A+1}\frac{3\varepsilon_B}{2\varepsilon_B+1}
\bm{G}(\mathbf{r}_A,\mathbf{r}_B,\omega)
\end{equation}
with $\varepsilon_{A,B}\equiv\varepsilon(\mathbf{r}_{A,B},\omega)$ and 
$\bm{G}(\mathbf{r}_A,\mathbf{r}_B,\omega)$ the uncorrected Green function. In 
the static limit, and noting that the excitons are embedded in the same crystal, 
we thus find that
\begin{equation}
\bm{\Gamma}_\mathrm{loc}(\mathbf{r}_A,\mathbf{r}_B) = \left( 
\frac{3\varepsilon_\mathrm{Cu_2O}}{2\varepsilon_\mathrm{Cu_2O}+1} \right)^2 
\bm{\Gamma}_\varepsilon(\mathbf{r}_A,\mathbf{r}_B)
\end{equation}
with the static bulk Green function inside a material with the permittivity 
$\varepsilon_\mathrm{Cu_2O}$
\begin{equation}
\bm{\Gamma}_\varepsilon(\mathbf{r}_A,\mathbf{r}_B) = 
\frac{1}{\varepsilon_\mathrm{Cu_2O}} 
\bm{\Gamma}_0(\mathbf{r}_A,\mathbf{r}_B).
\end{equation}
If we choose $\varepsilon_\mathrm{Cu_2O}=7.5$, this amounts to 
\begin{equation}
\bm{\Gamma}_\mathrm{loc}(\mathbf{r}_A,\mathbf{r}_B) \simeq 0.26 
\bm{\Gamma}_0(\mathbf{r}_A,\mathbf{r}_B).
\end{equation}

In the next step, we have to include the boundaries of the planar crystal 
cavity. In principle, one would need to construct the dyadic Green function for 
a planar three-layer system as done in Ref.~\cite{Tomas95} and then take its
static limit. Before we do that, we will estimate the effect of multiple 
scattering between the boundaries. The image of each dipole at one interface 
induces another image at the other interface and so on. By summing up the 
resulting geometric series, one finds an enhancement factor of
\begin{equation}
D_p\simeq\left[ 1-r_p^+ r_p^- \right]^{-1} \simeq 2.4
\end{equation}
if we choose $\varepsilon_\mathrm{Cu_2O}=7.5$, resulting in 
$r_p^\pm\simeq0.76$. It is obvious that this approach can only yield 
approximate results as the orientation of the dipoles inside the cavity are not 
taken into account.

We thus have to construct the zero-frequency limit of the known dyadic Green 
function for a planar three-layer system. Commonly, this is done using a 
two-dimensional Fourier (or Weyl) transform \cite{BuhmannI}
\begin{equation}
\bm{G}^{(S)}(\mathbf{r},\mathbf{r}',i\xi)=
\int d^2q\,e^{i\mathbf{q}\cdot(\mathbf{r}-\mathbf{r}')}
\bm{G}^{(S)}(\mathbf{q},z,z',i\xi)
\end{equation}
with 
\begin{gather}
\bm{G}^{(S)}(\mathbf{q},z,z',i\xi) = \frac{1}{8\pi^2b} \sum\limits_{\sigma=s,p}
\bigg\{ \frac{r_-^\sigma r_+^\sigma e^{-2bd}}{D^\sigma} \nonumber\\ \times 
\left[ \mathbf{e}^+_\sigma \otimes \mathbf{e}^+_\sigma e^{-b(z-z')} 
+\mathbf{e}^-_\sigma \otimes \mathbf{e}^-_\sigma e^{-2bd+b(z-z')}\right] 
\nonumber \\ +\frac{1}{D^\sigma} \left[ \mathbf{e}^+_\sigma \otimes 
\mathbf{e}^-_\sigma r_-^\sigma e^{-b(z+z')} +\mathbf{e}^-_\sigma \otimes 
\mathbf{e}^+_\sigma r_+^\sigma e^{-2bd+b(z+z')} \right]
\bigg\}
\end{gather}
as the Weyl components of the Green tensor. The $r_\pm^\sigma$ are the Fresnel
reflection coefficients at the upper and lower crystal-air interface,
respectively. The function $D^\sigma=1-r_-^\sigma r_+^\sigma e^{-2bd}$ accounts
for multiple reflections.

In this notation, $\mathbf{q}$ denotes the in-plane wavevector components and 
$b=\sqrt{\xi^2/c^2\varepsilon(i\xi)+q^2}$ its component normal to the planar 
surfaces. In view of the nonretarded limit, we can set $b\simeq q$. In the same 
limit, only the $p$-polarized waves will contribute, hence we set $\sigma=p$ 
with
\begin{gather}
\mathbf{e}_p^\pm=\mp \frac{b}{\kappa} \left[ \cos\varphi\mathbf{e}_x
+\sin\varphi\mathbf{e}_y \right] -\frac{iq}{\kappa} \mathbf{e}_z \nonumber\\
\simeq -\frac{q}{\kappa} \left[ 
\pm\cos\varphi\mathbf{e}_x\pm\sin\varphi\mathbf{e}_y+i\mathbf{e}_z \right].
\end{gather}
We restrict ourselves to the case in which the excitons are created at $z=z'$.
The coordinate system is oriented such that $\mathbf{r}-\mathbf{r}'$ points 
along the $x$-axis, hence 
$\mathbf{q}\cdot(\mathbf{r}-\mathbf{r}')=qx\cos\varphi$.

Expanding out the dyadic products of polarization unit vectors and using the 
integral representation of the Bessel functions, one finds
\begin{gather}
\int\limits_0^{2\pi}d\varphi\,e^{iqx\cos\varphi} \left[ 
\mathbf{e}_p^+\otimes\mathbf{e}_p^+ +\mathbf{e}_p^-\otimes\mathbf{e}_p^- 
\right] \nonumber\\
=\frac{2\pi q^2}{\kappa^2}
\begin{pmatrix}
J_0(qx)-J_2(qx) & 0 & 0 \\ 0 & J_0(qx)+J_2(qx) & 0 \\ 0 & 0 & -2J_0(qx)
\end{pmatrix}
\end{gather}
and
\begin{gather}
\int\limits_0^{2\pi}d\varphi\,e^{iqx\cos\varphi}
\mathbf{e}_p^\pm \otimes\mathbf{e}_p^\mp \nonumber\\
=-\frac{\pi q^2}{\kappa^2}
\begin{pmatrix}
J_0(qx)-J_2(qx) & 0 & \pm 2J_1(qx) \\
0 & J_0(qx)+J_2(qx) & 0 \\
\mp 2J_1(qx) & 0 & -2J_0(qx)
\end{pmatrix}.
\end{gather}
The remaining $q$-integrals can be performed by expanding the Bessel functions
into power series and using ($r_-=r_+\equiv r$)
\begin{equation}
\int\limits_0^\infty dq\,\frac{q^n}{1-r^2e^{-2qd}} = \frac{n!}{(2d)^{n+1}}
\operatorname{Li}_{n+1}(r^2)\,,
\end{equation}
\begin{equation}
\int\limits_0^\infty dq\,\frac{e^{-2qz}q^n}{1-r^2e^{-2qd}} =
\frac{n!}{(2d)^{n+1}} \Phi_{n+1}(r^2,z/d)\,,
\end{equation}
\begin{equation}
\int\limits_0^\infty dq\,\frac{e^{-2q(d-z)}q^n}{1-r^2e^{-2qd}} =
\frac{n!}{(2d)^{n+1}} \Phi_{n+1}(r^2,1-z/d)\,,
\end{equation}
where $\operatorname{Li}_n(x)$ is the polylogarithm function and $\Phi_n(x,z)$
the Lerch Phi function. These special functions are very well approximated by
\begin{equation}
\operatorname{Li}_n(r^2)\simeq r^2\,,\quad 
\Phi_n(r^2,z/d) \simeq (d/z)^n\,,\quad n\ge 3,
\end{equation}
with errors less than $2.5\%$. We can identify two types of integrals,
\begin{gather}
I_m=\int\limits_0^\infty dq\,\frac{r^2e^{-2qd}}{1-r^2e^{-2qd}} q^2 J_m(qx)
\simeq \frac{(m+2)!}{m!} \left( \frac{x}{4d} \right)^m \nonumber\\ \times
\frac{r^2}{(2d)^3}
{}_2F_1\left(\frac{m+3}{2},\frac{m+4}{2},m+1,-\frac{x^2}{4d^2}\right) \,,
\end{gather}
\begin{gather}
\tilde{I}_m=\int\limits_0^\infty dq\,\frac{re^{-2qz}}{1-r^2e^{-2qd}} q^2
J_m(qx) \simeq r \frac{(m+2)!}{m!} \left( \frac{x}{4z} \right)^m \nonumber\\ 
\times \frac{1}{(2d)^3}
{}_2F_1\left(\frac{m+3}{2},\frac{m+4}{2},m+1,-\frac{x^2}{4z^2}\right) \,,
\end{gather}
with the help of which we can express the dyadic Green function in the required
nonretarded limit. In the simplest case in which the excitons are excited
halfway between the plate interfaces, $z=d/2$, we find
\begin{gather}
4\pi\kappa^2\bm{G}^{(S)}(\mathbf{r},\mathbf{r}',i\xi) \nonumber\\
\simeq
\operatorname{diag}(I_0-I_2-\tilde{I}_0+\tilde{I}_2,
I_0+I_2-\tilde{I}_0-\tilde{I}_2,-2I_0-2\tilde{I}_0).
\end{gather}

\begin{figure}[ht]
\includegraphics[width=8cm]{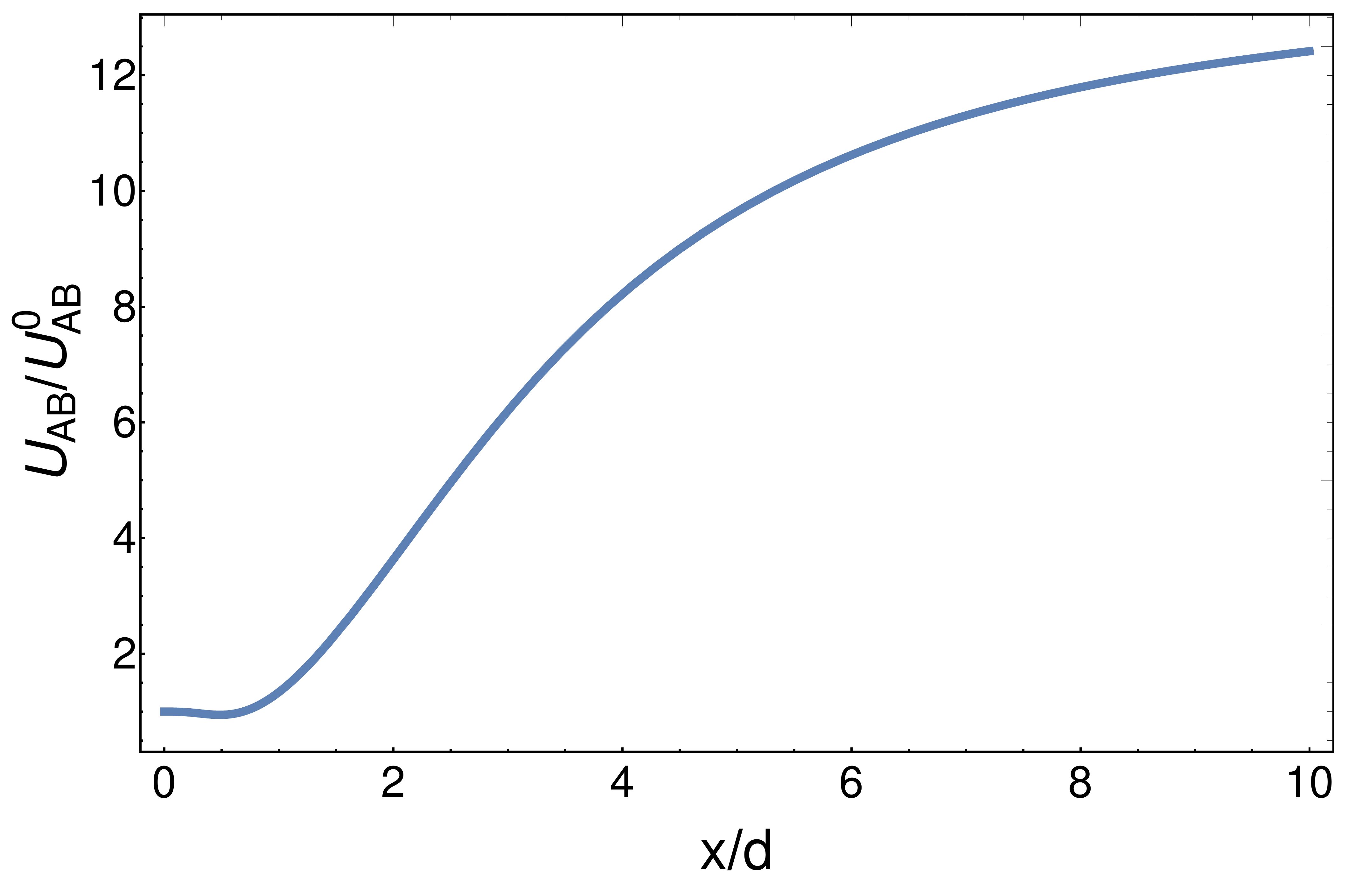}
\caption{\label{fig:3layer} Enhancement factor of the van der Waals interaction
potential between excitons separated by a distance $x$ inside a bulk
$\mathrm{Cu_2O}$ crystal with thickness $d$.}
\end{figure}

Inserted into the expression for the van der Waals potential and normalizing
with respect to the local-field corrected bulk value, we find the enhancement
factor due to the presence of the boundaries as shown in Fig.~\ref{fig:3layer}.
There we show the enhancement of the nonretarded van der Waals interaction
between two excitons located inside a crystal of thickness $d$ at a vertical
position $z=d/2$, and separated by a horizontal distance $x$. Note that, due to
the lack of any other length scale in the considered nonretarded limit, the
figure can be scaled to arbitrary values of $x$ and $d$, as they are the only
remaining length scales in the system. We see that, depending on the ratio
$x/d$, the enhancement of the van der Waals potential due to the restricted
geometry can easily exceed factors of 5 or more. 

This explains why the theoretical values of the Rydberg blockade efficiency
given in Ref.~\cite{Kazimierczuk14} differ substantially from the data extracted
from the experiment. The crystal thickness was given as $d=34\mu$m, and the
Rydberg blockade radius estimated to be around $30\mu$m. The enhancement factor
can therefore estimated to be around 2, which accounts for the discrepancy
between theory and experiment in Ref.~\cite{Kazimierczuk14}.

\section{Rydberg atoms in front of a perfectly conducting plate}
\label{sec:atoms}

We will now apply the concepts described above to the van der Waals interaction 
of Rydberg atoms close to a perfect mirror, which can be viewed as a model 
system for a superconducting strip line cavity \cite{Hattermann17} or similar
surfaces \cite{Thiele14,WeissFortagh2015}. Two modifications are required to 
the simple perturbative model. First, nondegenerate perturbation theory fails 
when describing interactions between two Rydberg states \cite{Stanojevic08} as 
the F\"orster defect \cite{Saffman08} can become very small for resonant 
interactions. Hence, it becomes necessary to exactly diagonalize the van der 
Waals interaction Hamiltonian. Second, the spatial extent of the wavefunction 
of a Rydberg atom requires the inclusion of higher-order multipole moments of 
the interaction \cite{Crosse10}.

Including higher-order terms up to quadrupole-quadrupole interaction, the 
interaction Hamiltonian then reads
\begin{equation}
\begin{aligned}
 \hat{H}_{int} &= 
\hat{\mathbf{d}}_A\cdot\bm{\Gamma}_0(\mathbf{r}_A,\mathbf{r}_B)
\cdot\hat{\mathbf{d}}_B \\ 
&+ \hat{\mathbf{d}}_A\cdot\bm{\Gamma}_0(\mathbf{r}_A,\mathbf{r}_B)
\overleftarrow{\nabla}_B : \hat{\bm{Q}}_B\\
&+ \hat{\bm{Q}}_A : \overrightarrow{\nabla}_A 
\bm{\Gamma}_0(\mathbf{r}_A,\mathbf{r}_B)\cdot\hat{\mathbf{d}}_B \\
&+ \hat{\bm{Q}}_A:\overrightarrow{\nabla}_A \bm{\Gamma}_0({\bf 
r}_A,{\bf r}_B)\overleftarrow{\nabla}_B : \hat{\bm{Q}}_B
\end{aligned}\label{eq:H_int}
\end{equation}
where indices and arrows on the gradient operator indicate the spatial 
coordinate and the tensor index with respect to which the differentiation is 
performed. Here, the (tensorial) quadrupole moment operators are denoted by 
$\hat{\bm{Q}}_{A,B}$, and the notation $:$ indicates the contraction of two 
indices (Hadamard product).

In order to calculate the gradient of the static Green function for a perfect 
mirror, Eq.~\eqref{eq:staticGmirror}, it is convenient to define a matrix 
$\bm{A}=\overrightarrow{\nabla}\frac{\bm{r}}{r}$ with 
$\bm{r}=\mathbf{r}_A-\mathbf{r}_B$. For the free-space part 
$\bm{\Gamma}_0^{\mathrm{fs}}$, we can then write
\begin{equation}
 \begin{aligned}
  \overrightarrow{\nabla}_A\bm{\Gamma}_0^{\mathrm{fs}}(\bm{r},\omega) = 
\frac{1}{4\pi}&\left\{\frac{3}{r^4}\mathbf{e}_r\otimes\bm{I} - 
\frac{9}{r^4}\mathbf{e}_r\otimes\mathbf{e}_r\otimes\mathbf{e}_r
\right. \\
 + &\left.\frac{3}{r^3}\left(\bm{A}\otimes\mathbf{e}_r + 
A_{ik}e_{r_j}\right)\right\}
 \end{aligned} \label{eq:qgd}
\end{equation}
and
\begin{equation}
 \begin{aligned}
  \bm{\Gamma}_0^{\mathrm{fs}}(\bm{r},\omega)\overleftarrow{\nabla}_B = \frac{c^2}{4\pi 
\omega^2}&\left\{-\frac{3}{r^4}\bm{I}\otimes\mathbf{e}_r + 
\frac{9}{r^4}\mathbf{e}_r\otimes\mathbf{e}_r\otimes\mathbf{e}_r 
\right. \\
 + &\left.\frac{3}{r^3}\left(\mathbf{e}_r\otimes\bm{A} + 
A_{ik}e_{r_j}\right)\right\}.
 \end{aligned} \label{eq:dgq}
\end{equation}
This means that the double gradient of the nonretarded Green function required
for the quadrupole-quadrupole interaction becomes
\begin{equation}
 \begin{aligned}
&\overrightarrow{\nabla}_A\bm{\Gamma}_0^{\mathrm{fs}}(\bm{r},\omega)\overleftarrow{\nabla}_B = 
\frac{c^2}{4\pi \omega^2}\\
&\times\left\{\frac{12}{r^5}\mathbf{e}_r\otimes\bm{I}\otimes\mathbf{e}
_r - 
\frac{36}{r^5}\mathbf{e}_r\otimes\mathbf{e}_r\otimes\mathbf{e}
_r\otimes\mathbf{e}_r \right. \\
 &+ \left.\frac{3}{r^4}\left(3A_{ij}e_{r_k}e_{r_l} + 
3A_{ik}e_{r_j}e_{r_l} + 3A_{il}e_{r_j}e_{r_k} - 
A_{il}\delta_{jk}\right)\right. \\
&+ \left. \frac{3}{r^3}\left(A_{ik}A_{jl} + A_{ij}A_{kl} - 
3e_{r_i}A_{jl}e_{r_k} - 3e_{r_i}e_{r_j}A_{kl}\right)\right\}.
 \end{aligned}\label{eq:qgq}
\end{equation}
The scattering part of the tensor, corresponding to the second line
in Eq.~\eqref{eq:staticGmirror}, is calculated accordingly. 

The interaction Hamiltonian is then expanded into a basis spanned by the states 
$|\psi\rangle = |n,l,j,m_j\rangle$ where $n$ is the principal quantum number, 
$l$ is the angular momentum quantum number, the total angular momentum $j = 
l\pm\frac{1}{2}$ accounts for the spin, and $m_j \in [-j,j]$ is the projection 
of $j$ onto the quantization axis. The binding energies are determined via the
quantum defect $\delta_{nlj}$ and the modified Rydberg series
\begin{equation}
 E_{n,l,j} = -\frac{E_{\mathrm{Ry}}}{(n-\delta_{nlj})^2}.
\end{equation}
Quantum defects for rubidium Rydberg states with $n>11$ were calculated in 
Refs.~\cite{LiMourachko03,Han06}, whereas for $n<11$, data can be found 
in Ref.~\cite{Sansonetti06}. Detailed measurements of absolute excitation 
energies with extracted quantum defects can be found in Ref.~\cite{Mack11}.

The quadrupole operator is defined as 
$\hat{\bm{Q}}=\frac{q}{2}\hat{\mathbf{r}}\otimes\hat{\mathbf{r}}$ and can be 
expressed in terms of spherical harmonics \cite{Crosse10}. The radial part of 
both $\hat{\mathbf{d}}$ and $\hat{\bm{Q}}$, i.e. the matrix elements $\langle 
n,l,j|\hat{r}|n',l',j'\rangle$ and $\langle n,l,j|\hat{r}^2|n',l',j'\rangle$, 
respectively, are calculated numerically using Numerov's method integrating 
inward until an inner cutoff point. 
For the angular components, we use the Wigner--Eckart theorem to write 
\begin{equation}
\begin{aligned}
&\langle l,j,m_j |Y_{kq}|l',j',m_j'\rangle = (-1)^{j+j'-m_j+k+\frac{1}{2}}\\
&\times\sqrt{\frac{(2j+1)(2j'+1)(2l+1)(2l'+1)(2k+1)}{4\pi}}\\
&\times\tj{j}{k}{j'}{-m}{q}{m'}\tj{l}{k}{l'}{0}{0}{0}\sj{l}{j}{\frac{1}{2}}{j'}{l'}{k},
\end{aligned}
\end{equation}
where we set $s = s' = \frac{1}{2}$ and use the Wigner-$3j$-symbol 
$\tj{l}{l'}{L}{m}{m'}{M}$ as well as the Wigner-$6j$-symbol 
$\sj{l}{j}{s}{j'}{l'}{k}$.

Now we turn to a setup of two atoms in parallel alignment in front of a 
perfect mirror, see Fig.~\ref{fig:2atome} with $z_- = 0$. In free space, due to isotropy
one can always choose the quantization axis of the atoms to be parallel to the 
the intermolecular axis connecting both atoms. This implies a conservation of 
the total projection of the angular momentum $M=m_A+m_B$. Thus, the interaction 
Hamiltonian decomposes into blocks of constant $M$~\cite{Saffman08} which 
simplifies the computation considerably. 

The presence of a macroscopic body breaks that isotropy and hence the 
conservation of $M=m_A+m_B$. Therefore, one has to include a far larger basis 
set for the full diagonalization. In Ref.~\cite{Stanojevic08}, the authors 
considered two Rb atoms in states $70p+70p$, and focussed on very few strongly 
dipole- and quadrupole-coupled asymptotes for the diagonalization, resulting in 
a basis set of 142 two-atom states, i.e. a $142\times142$ matrix that had to be 
diagonalized. Breaking isotropy by introducing a macroscopic body immediately 
generates a $700\times700$-matrix. In general, convergence for 
free-space dipole-dipole interaction of Rydberg atoms is often only achieved 
with a basis set of 4000 to 6000 states \cite{Cabral2011}.

\begin{figure}
 \includegraphics[width=\columnwidth]{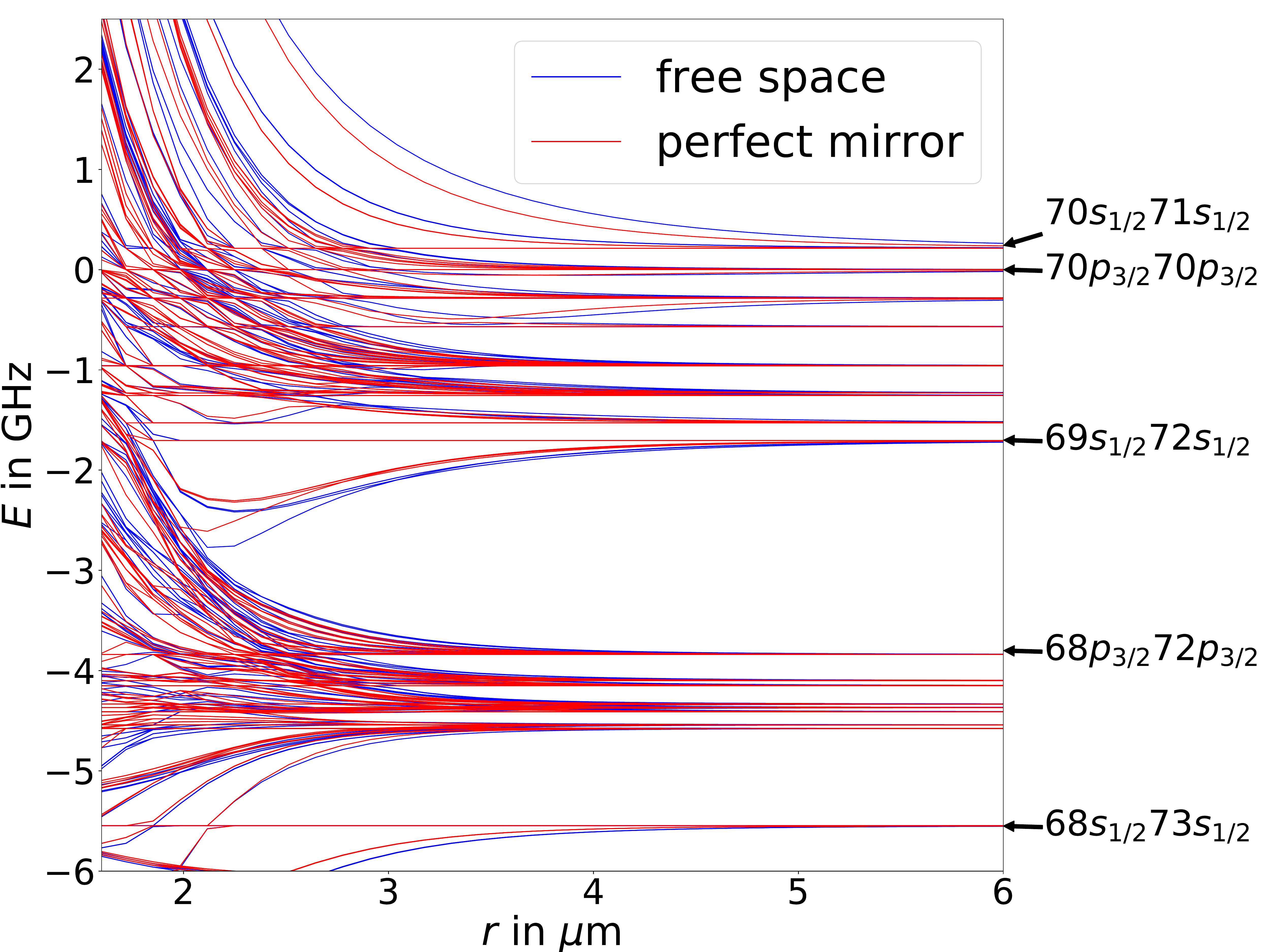}
\caption{We depict the interaction potential of several states around the 
Rubidium $70p_{3/2}70p_{3/2}$ asymptote at $0\,\mathrm{GHz}$. Blue lines: 
free-space interaction. Red lines: interaction of two atoms in parallel 
alignment at a distance $d \approx1\,\mu\mathrm{m}$ in front of a perfect 
mirror.}
\label{fig:Rydberg}
\end{figure}

In Fig.~\ref{fig:Rydberg}, we present an example of the effects of a nearby 
surface on the van der Waals interaction between two Rydberg atoms. Here, we 
take into account the same basis set as in Ref.~\cite{Stanojevic08}, including 
hyperfine splitting for $d$- and $f$-states. We use the the \textit{gerade} 
symmetrization, i.e. $|A,B\rangle \rightarrow\frac{1}{\sqrt{2}}\left( 
|A,B\rangle - (-1)^{l_A+l_B}|B,A\rangle\right)$ for a two-atom state 
$|A,B\rangle = |n_Al_Aj_Am_{j,A};n_Bl_Bj_Bm_{j,B}\rangle$. We show the 
interaction of two Rb atoms in free space (blue) and at a distance of 
$\sim1~\mu$m from a perfectly reflecting mirror (red). One observes notable 
changes in the interaction potential for interatomic distances 
$r<3.5\,\mu\mathrm{m}$. 

\begin{figure}
 \includegraphics[width=\columnwidth]{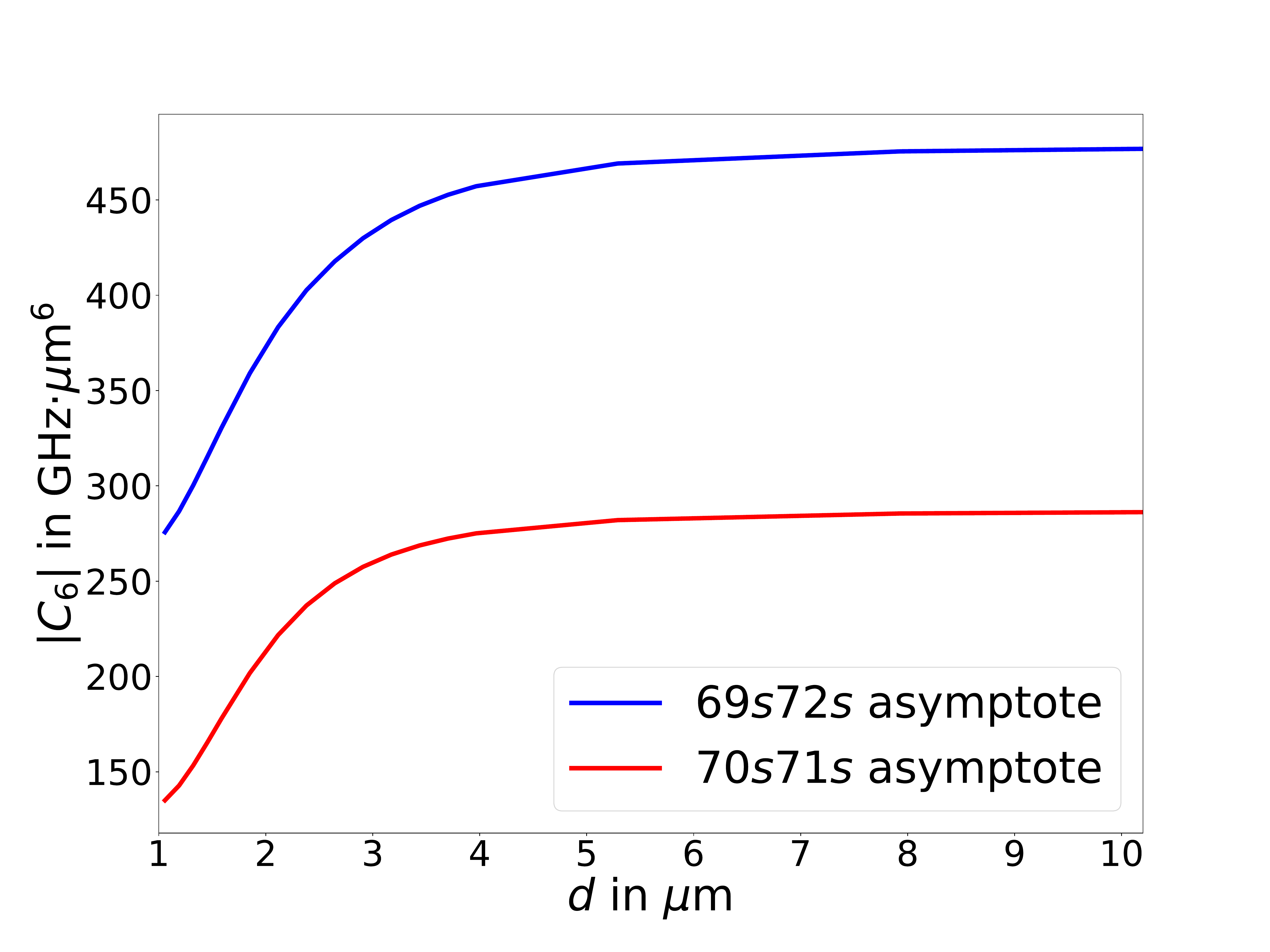}
 \caption{Development of the $C_6$ coefficient over distance $d$ between the atoms and the perfect mirror.}
 \label{fig:c6coefficients}
\end{figure}

As illustrative examples, We will investigate the $70s_{1/2}71s_{1/2}$ and 
$69s_{1/2}72s_{1/2}$ asymptotes further. 
Both potential curves are fitted to a $C_6/r^6$ potential as expected from the nonretarded van der Waals interaction. 
This can only be a rough estimate since we also consider multipole interactions up to quadrupole-quadrupole that lead to additional terms in the potential.
We find a general decrease in the $C_6$ coefficient with smaller distance $d$ between the atoms and the perfect mirror.
Results are shown in Figure~\ref{fig:c6coefficients}.
For the $70s_{1/2}71s_{1/2}$ asymptote, the free space coefficient is approximately $C_{6}^{\mathrm{fs}} \approx 287\,\mathrm{GHz}\cdot\mu\mathrm{m}^6$ while at a distance $d\approx 1\,\mu\mathrm{m}$ it reduces to $C_{6}^{\mathrm{mirror}}(d\approx1\,\mu\mathrm{m}) \approx 152\,\mathrm{GHz}\cdot\mu\mathrm{m}^6$.
For the $69s_{1/2}72s_{1/2}$ asymptote, we can extract $C_{6}^{\mathrm{fs}} \approx 478\,\mathrm{GHz}\cdot\mu\mathrm{m}^6$ and  $C_{6}^{\mathrm{mirror}}(d\approx1\,\mu\mathrm{m}) \approx 283\,\mathrm{GHz}\cdot\mu\mathrm{m}^6$.

\section{Conclusions}

The quantum electrodynamic approach to van der Waals interactions provides one 
with the notion of an exchange of virtual photons as the cause of that 
fluctuation-induced interaction. For atoms in excited states, the exchange 
of real photons contributes, too. We have shown here that, in the nonretarded 
limit, the contributions from virtual and real photons, i.e. the nonresonant 
and resonant contributions to the van der Waals potential, overcompensate one 
another to the extent that a distinction between them looks entirely 
artificial. However, it should be stressed that this is only true in the 
nonretarded limit in which the spatial dependence that resides in the dyadic 
Green function, factorizes from the frequency dependence.

The immediate result is that, perhaps unsurprisingly, in this limit, the 
quantum electrodynamic calculation coincides with the static dipole-dipole 
interaction commonly used in atomic Rydberg physics. However, the more general 
quantum electrodynamic approach provides one with the additional insight how 
macroscopic bodies alter the van der Waals interaction between highly excited 
atoms. For Rydberg atoms, it is easy to envisage situations in which the 
body-induced modification of the van der Waals potential can be significant. 
Indeed, as soon as the atom-surface distance becomes comparable to the 
interatomic separation, one has to expect contributions to the interaction 
potential that arises from the reflection of the photons off the body surface.

We have shown how to construct the van der Waals potential near a planar 
interface as well as inside a planar waveguide, and given an estimate of the 
enhancement factor of the van der Waals interaction of excitons in a thin 
crystal. Note, however, that in this specific example we only investigated the 
perturbative interaction that does not take degeneracies into account. Also, we 
have disregarded the absolute strength of the interaction which would require a 
detailed description of the transition dipole moments of the exciton states. 
This will be the subject of future work.

For Rydberg atoms near planar interfaces such as superconducting striplines, 
the direct diagonalization approach is more involved than in free space as the 
presence of the interface breaks the isotropy of space. This implies that the 
interaction Hamiltonian does no longer decompose into blocks with constant 
total magnetic quantum number as quantization axis and intermolecular axis no 
longer necessarily coincide. We have given an example of an Rb dimer 
horizontally aligned to a perfect conductor to illustrate the effect of the 
interface on the van der Waals potential and subsequently on the Rydberg 
blockade radius. With this, it is now possible to use macroscopic bodies as 
additional tuning devices for controlling interactions of Rydberg systems.

\acknowledgments We gratefully acknowledge discussions with R.~Schmidt. This 
work was partially supported by the Collaborative Research Centre SFB 652/3 
'Strong correlations in the radiation field' funded by the Deutsche 
Forschungsgemeinschaft and the Landesgraduiertenf\"orderung Mecklenburg-Vorpommern.

\end{document}